# Citizen Science Astronomy with a Network of Small Telescope: The Launch and Deployment of JWST

Lambert R. A.[a], Marchis, F.[a,b,*], Asencio, J.[b], Blaclard, G.[b], Sgro, L.A.[a], Giorgini, J.D.[c], Plavchan, P.[d], White, T.[e], Verveen, A.[e,] Goto, T.[e,] Kuossari, P.[e], Nagendra, S.[e], Loose, M.A.[e], Will, S.[e], Sibbernsen, K.[e], Pickering, J.W[e], Randolph, J.[e], Fukui, K.[e], Huet, P.[e], Guillet, B.[e], Clerget, O.[e], Stahl, S.[e], Yoblonsky, N.[e], Lauvernier, M.[e], Matsumura, T.[e], Yamato, M.[e], Laugier, J.M.[e], Brodt-Vilain, O.[e], Espudo, A.[e], Kukita, R.[e], Iida, S.[e], Kardel, S.[e], Green, D.[e], Tikkanen, P.[e], Douvas, A.[e], Billiani, M.[e], Knight, G.[e], Ryno, M.[e], Simard, G.[e], Knight, R.[e], Primm, M.[e], Wildhagen, B.[e], Poncet, J.[e], Frachon, T.[e], Shimizu, M.[e], Jackson, A.[e], Parker, B.[e], Redfern, G.[e], Nikiforov, P.[e], Friday, E.[e], Lincoln, K.[e], Sweitzer, J.[e], Mitsuoka, R.[e], Cabral, K.[e], Katterfeld, A.[e], Fairfax, M.[e]

[a]SETI Institute, 339 Bernardo Ave, Suite 200, Mountain View, CA, USA 94043; [b]Unistellar, 5 allée Marcel Leclerc,. bâtiment B,. 13008 Marseille, France; [c]Solar System Dynamics Group, Jet Propulsion Laboratory, California Institute of Technology, 4800 Oak Grove Drive, Pasadena, CA 91109, USA; [d]George Mason University, 4400 University Drive Fairfax, VA, 22030, USA; [e]Unistellar Citizen Scientist, Earth.

## ABSTRACT

We present a coordinated campaign of observations to monitor the brightness of the James Webb Space Telescope (JWST) as it travels toward the second Earth-Sun Lagrange point and unfolds using the network ofUnistellar digital telescopes. Those observations collected by citizen astronomers across the world allowed us to detect specific phases such as the separation from the booster, glare due to a change of orientation after a maneuver, the unfurling of the sunshield, and deployment of the primary mirror. After deployment of the sunshield on January 6 2022, the 6-h lightcurve has a significant amplitude and shows small variations that we cannot explain. These variations could be due to the deployment of the primary mirror or some changes in orientation of the space telescope. This work illustrates the power of a worldwide array of small telescopes, operated by citizen astronomers, to conduct large scientific campaigns over a long timeframe. In the future, our network and others will continue to monitor JWST to detect potential degradations to the space environment by comparing the evolution of the lightcurve.

**Keywords:** JWST, telescope network, digital telescope, citizen science, new astronomy.

*fmarchis@seti.org; phone +1 510 599 0604

## 1. INTRODUCTION

The Unistellar network, composed of 10,000 digital, smart and easy-to-use telescopes distributed worldwide was used to follow the James Webb Space Telescope (JWST) on its way to the Lagrange point from December 2021 to March 2022. The motivation of this campaign, led and organized by researchers at the SETI Institute, is twofold:

1. Provide accurate photometric and astrometric positions of the space telescope to monitor its trajectory as well as detect its deployment.

2. Allow our citizen scientists to witness in real time the activity of JWST during the separation from its booster, the deployment of its sunshield, and the opening of its primary mirror.

The success of the campaign has relied on the willingness of our community to observe the space telescope using our ephemeris tool and share their observations. The data processing and analysis has been conducted by the SETI

Institute and reports were sent to our citizen scientists. We describe below the Unistellar network, ephemeris tools, and our communication strategy to involve as many observers as possible in following JWST for three months.

### 1.1 Observations from the Unistellar Network

We present observations of JWST, mainly collected by the Unistellar network of citizen astronomers using their digital telescopes. The Unistellar telescopes[1,2], named *eVscope, eQuinox, eVscope 2*, are capable of aligning themselves and acquiring targets. In the case of JWST, we developed a specific web site described below to generate the ephemeris of the space telescope so citizen astronomers around the world could observe it.

The Unistellar observations were taken over the course of ~4 months, from launch on December 25 2021 all the way to JWST's final destination, the second Earth-Sun Lagrange point (L2) on March 18 2022. We received more than 145 observations from 55 observers located across North America, Japan, Europe, Russia, and Oceania and we have analyzed 35% of them so far. From these observations, we have extracted the astrometric positions, the photometric measurements and a partial lightcurve.

### 1.2 Ephemeris Tools

A moving target ephemeris webpage was designed to aid users by calculating the celestial coordinates (right ascension and declination) of JWST at a given time from a given location. Users navigating to the webpage from their smartphone were able to choose JWST from a list of available targets, then input their location and time of observation, from which a list of astrometric positions is displayed if JWST is visible for them. This page also generates a deeplink, which automatically opens the Unistellar application that controls the eVscope and directly inputs the target right ascension, declination, gain, and exposure time, allowing the user to begin observing right away if they own a Unistellar telescope.

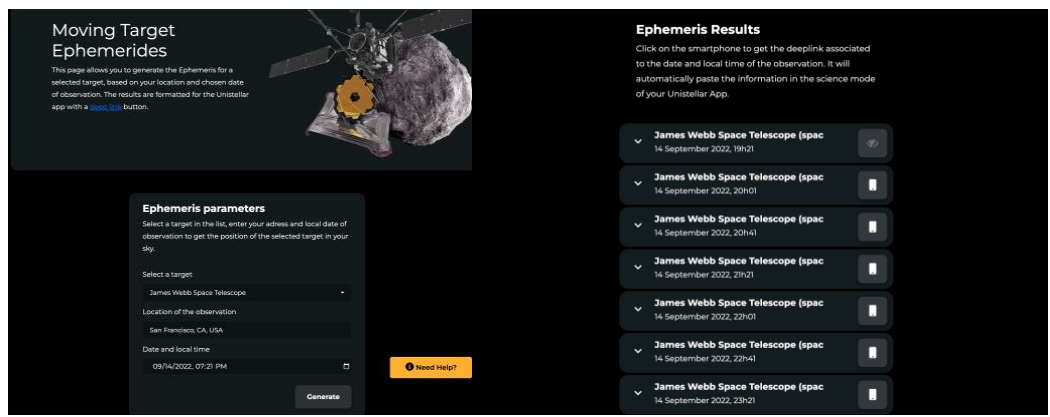

Figure 1. Ephemeris page available on the Unistellar web site (https://unistellaroptics.com/ephemeris/) allowing users to generate the positions of human-made objects like JWST. The page automatically generates the position at the proper cadence for the eVscope and filters out an observation if the object is too low in elevation. A deeplink is also generated (indicated by the phone icon), which automatically opens the Unistellar App and allows the user to point their eVscope to the target.

The web page itself uses an API to query JPL Horizons[3] with a target, observing location, and time to construct an ephemeris for the observer. The code determines the mean velocity of the target and suggests an observing session length based on how long the target will remain in the field of view of the eVscope (typically half degree).

## 2. OBSERVATIONS

### 2.1 Lightcurves Extracted from Photometry

All observations were taken as a series of 4-second exposures. Since observations were taken by different observers in different environments (from cities with Bortle 4-5 sky to the countryside with Bortle 2 sky), the duration of observations was variable. Two-minutes dark frames were taken at the end of every observation by putting the lens cap on the eVscope and recording images. The data recorded by the eVscope was then uploaded to the Unistellar servers for processing.

Image processing was done using a planetary defense pipeline capable of dark subtracting, plate-solving, and stacking the uploaded images. Master dark frames were made by a median combination and these master darks were subtracted from each image of the field. The resulting dark-subtracted frames were then plate-solved using an API that coordinated with astrometry.net. Sets of images in which a WCS solution was found were then stacked based on the mean relative velocity with which JWST was moving at the time. JWST was identified using the predicted coordinates from the JPL Horizons ephemeris and confirmed manually for each data set. The magnitude of JWST was estimated using aperture photometry on each stacked image, comparing the telescope to reference stars in the Gaia database[4]. Figure 2-5 are a representative sample of the extracted lightcurves in chronological order that we have obtained from the data so far, corresponding to 36 observing sets. Of the 145 observations submitted to Unistellar, 53 had their astrometric positions of JWST submitted to the Minor Planet Center.

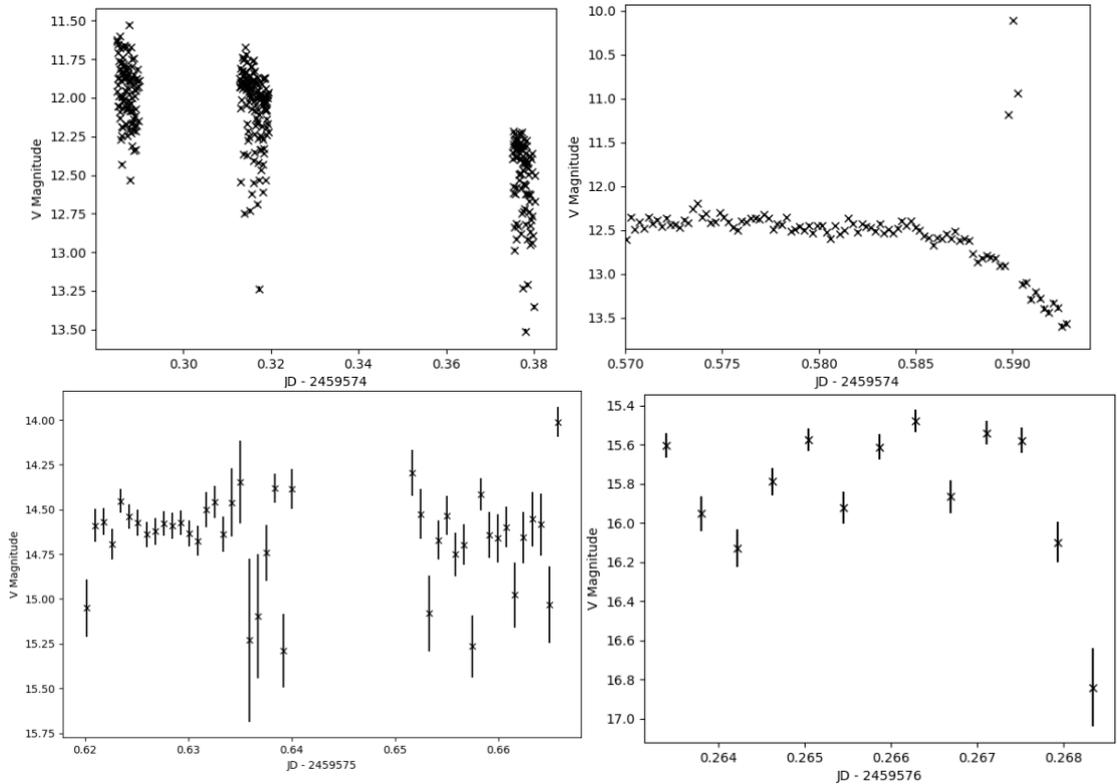

Figure 2. JWST lightcurves recorded from 2021-12-25 ~18:40 UT to 2021-12-27 18:30 UT, or 6h and 2 days after the launch, respectively. The brightness of the space telescope drops from ~12 to 16.2 magnitude. During this period of time, three Mid-Course Correction (MCC) maneuvers were applied to achieve a proper Sun-Earth/Moon L2 Libration point orbit.

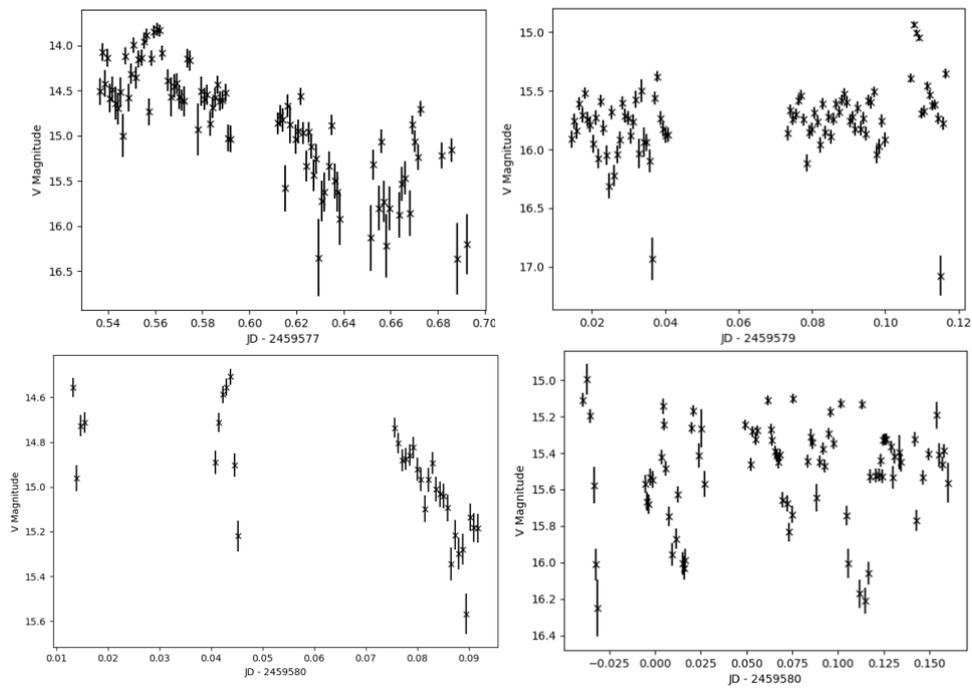

Figure 3. JWST lightcurves recorded from 2021-12-29 ~00:00 UT to 2021-12-31 15:40 UT. Over 5h of observations, the amplitude of the lightcurve reaches 2.5 magnitude, suggesting specular reflection on the surface of the shiny space telescope.

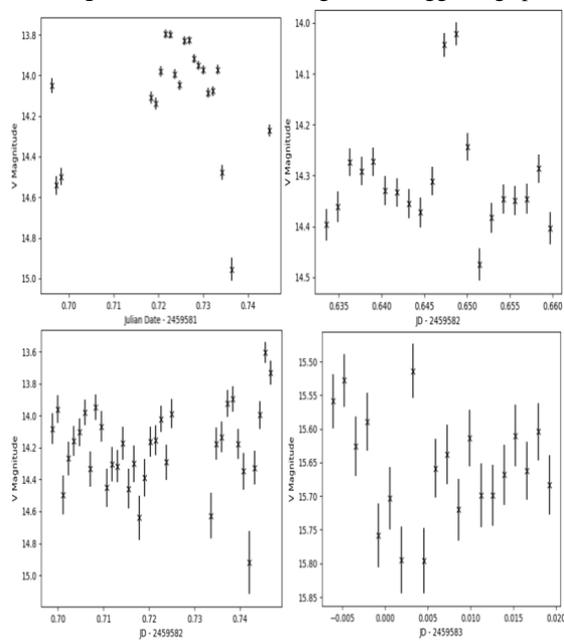

Figure 4. JWST lightcurves recorded from 2022-01-02 ~04:40 UT to 2022-01-03 ~12:30 UT.

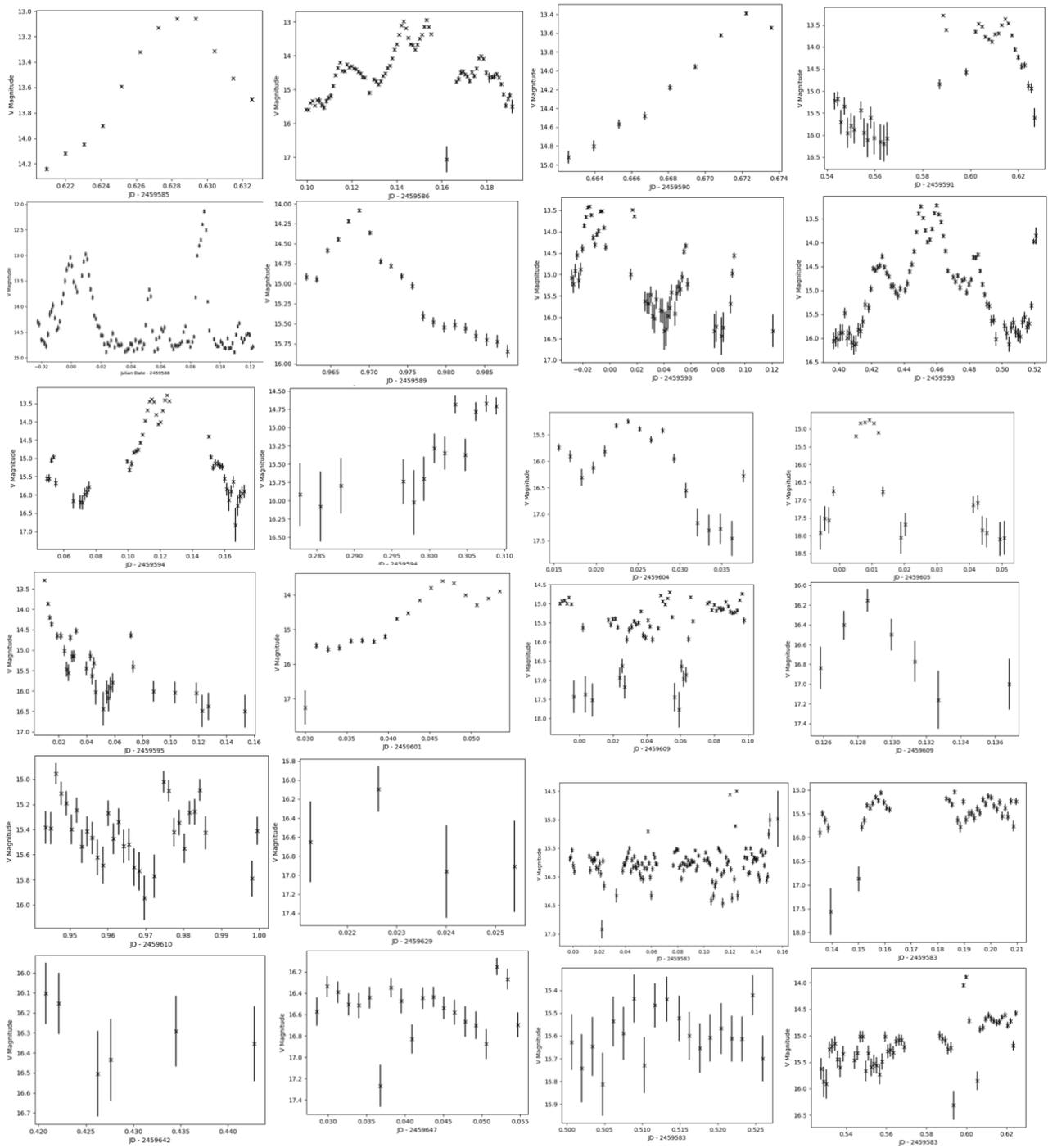

Figure 5. JWST lightcurves recorded from 2022-01-04 ~00:00 UT to 2022-03-08. On 2022-01-06 (JD 2459586) the first instance of a double peak within the lightcurve is visible.

## 3. RESULTS

### 3.1 Post-Mid-Course-Correction-Maneuver Flare

On December 26 2021, citizen astronomer Kendra Sibbernsen, located in Nebraska, USA, observed JWST with her eVscope shortly after the third Mid-Course Correction (MCC) maneuvers. Figure 6 shows that the spacecraft brightness was relatively stable for ~0.5h with a magnitude of 12.5 in the V band. The spacecraft brightness increased by 2.5 magnitude and reached a magnitude of 10.109 +/- 0.008, peaking at 2021-12-26 02:09:39.06 UT and lasting around 20 s. We have no direct information on the orientation of JWST at this time, but shortly after the separation from the booster, when the solar array had just been deployed, a camera on board the booster clearly showed an increase in brightness due to specular reflection on the solar array (Video 1). It is possible that after the MCC maneuver, the orientation of the spacecraft had changed positioning of the solar panel so a specular reflection could be seen from Earth.

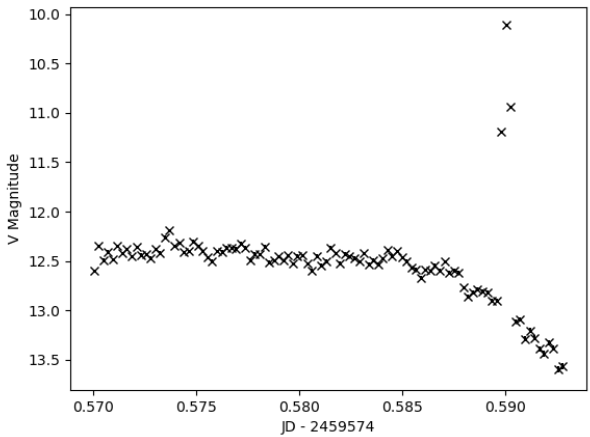

Figure 6. JWST lightcurve recorded on December 26 2021 by citizen astronomer Kendra Sibbernsen. At 02:09:39.06 UT, the brightness of the space telescope increased by 2.5 magnitude for 20s. The origin of this specular reflection is unknown since the orientation of the spacecraft was not provided to us.

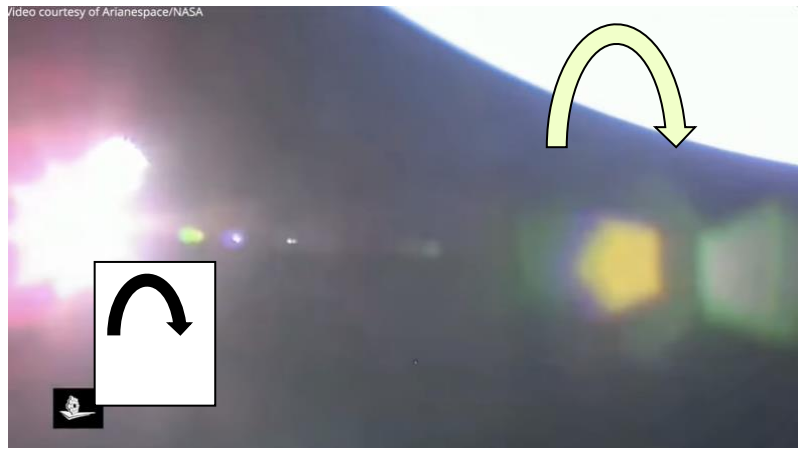

Video 1. Separation of JWST from its booster and deployment of the solar array observed from the booster. A sudden increase of brightness was recorded. The flare reported in Figure 5 shortly after an MCC maneuver could be due to the same type of specular reflection (video courtesy ArianeEspace/NASA): http://dx.doi.org/doi.number.goes.here

## 3.2 Observation of Booster Rocket After Detachment

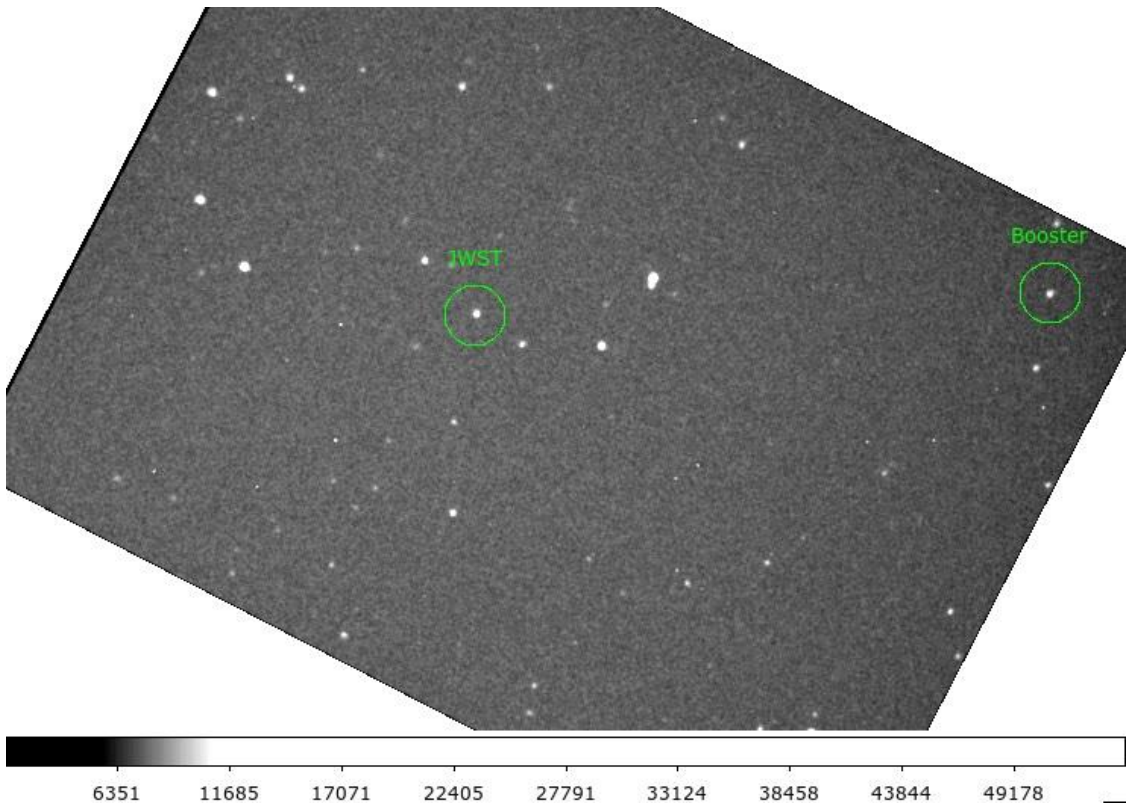

Video 2. Observation of the undeployed JWST and its booster shortly after the separation by citizen scientist Aad Verveen from the Netherlands : http://dx.doi.org/doi.number.goes.here

The booster rocket was observed following JWST approximately 7 hours after separation occurred. At the beginning of the observation, the booster was an estimated 3028 ± 5 km from JWST. Over the course of six minutes, the booster fell ~30 km further behind JWST, drifting with a relative speed of 300 km/h. At the time of the observations, the booster had an estimated magnitude of 12.43 ± 0.02 with no detectable change in magnitude for the duration of the observation.

Table 2. The difference in the RA and Dec positions of JWST and the booster rocket over the course of six minutes. Positions were measured with SAOImage DS9[5] using the plate-solved images.

| Time (UT) | Difference in RA (°) [± 0.0005°] | Difference in Dec (°) [± 0.0005°] |
| --- | --- | --- |
| 2021-12-25 18:50:37 | 0.3640 | -0.0120 |
| 2021-12-25 18:51:21 | 0.3650 | -0.0130 |
| 2021-12-25 18:52:05 | 0.3650 | -0.0120 |
| 2021-12-25 18:52:48 | 0.3660 | -0.0130 |
| 2021-12-25 18:53:32 | 0.3670 | -0.0120 |
| 2021-12-25 18:54:12 | 0.3670 | -0.0130 |
| 2021-12-25 18:54:55 | 0.3680 | -0.0120 |
| 2021-12-25 18:55:39 | 0.3680 | -0.0120 |
| 2021-12-25 18:56:23 | 0.3680 | -0.0130 |

### 3.3 Global Brightness Variation

Over the course of the campaign, the overall brightness of JWST appeared to vary based on its distance to the observer (from 0 to 0.01 AU) and the projected size of the space telescope. In Figure 7, we divide this variation into 4 major phases, depicted as yellow lines highlighting the overall trend in brightness.

1. Phase 1 is characterized by a sharp decrease in brightness as JWST drifted away from the Earth from 0 to ~430,000 km. During this time, the sunshield has not been deployed and the distance/projected size of the space telescope is the dominant factor in the drop of brightness. The orientation of the spacecraft has remained the same after the last MCC maneuver.
2. During phase 2, the sunshield begins deployment. During this time, the brightness increases due to the increased projected size and surface brightness from the reflective sunshield. Observations also show significantly more variation in the lightcurve, with for instance the first detections of peaked structure in the lightcurve on January 6 2022.
3. After the space telescope is fully unfurled, phase 3 begins. At this point the projected area is no longer increasing and JWST begins to dim once again as it travels to $L_2$.
4. The final phase begins once JWST reaches its final destination, $L_2$. The distance and projected size remain largely static (V~16.5). The brightness of the space telescope varies slightly between observations but shows no global trend towards dimming or brightening. The last observation shown here was taken on March 19th from the UK by citizen scientist Mark Fairfax.

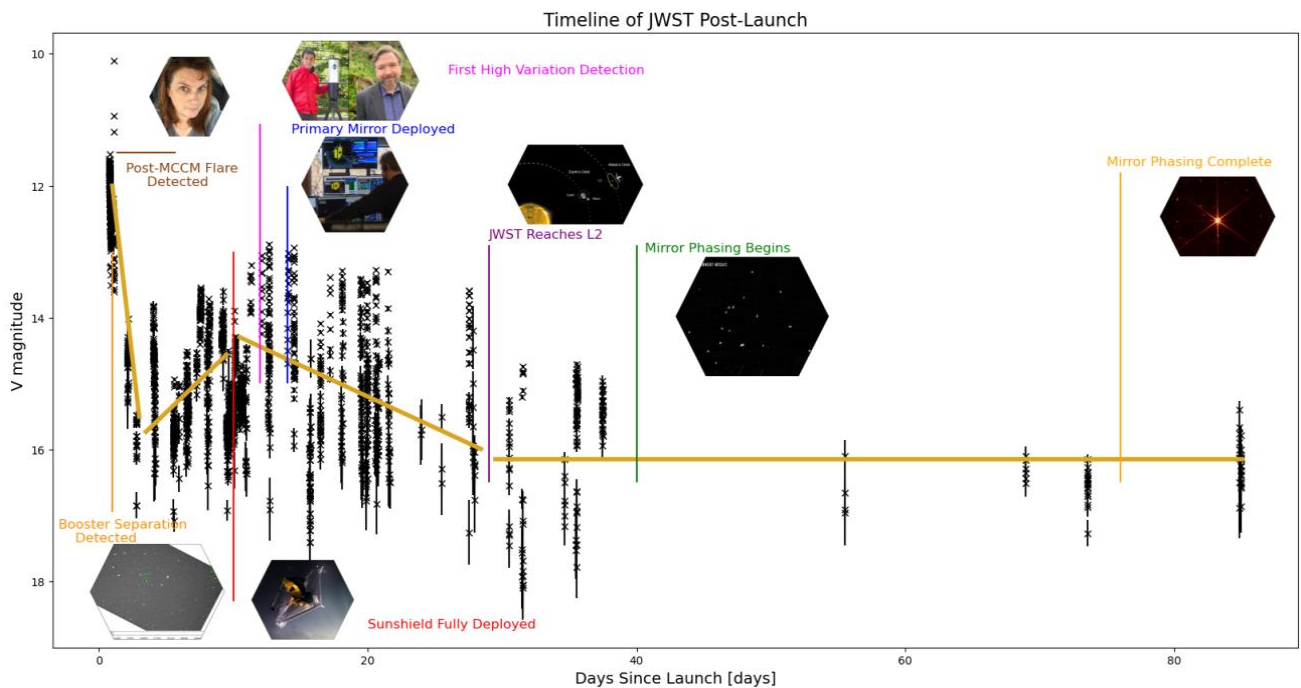

Figure 7. Global light curve variation showing the changes in brightness. Yellow lines indicate the overall trend in the brightness of JWST at various stages of its deployment.

### 3.4 Disappearance of Two Peaks in the Lightcurve on January 8 2022

In our investigations of the data, we compared the lightcurves extracted from data obtained by citizen scientists with a full, ~6-hour period lightcurve extracted from an observation taken by Peter Plavchan on January 12 2022 using the

George Mason University 0.8m telescope[6]. This data was dark subtracted and flat-fielded by using calibration observations taken that same night. Observations were aligned using the image analysis software, AstroimageJ[7]. We used aperture photometry to extract the lightcurve from this data set (see Figure 8) so that we might compare it to our observations. The origin of this ~6-hour lightcurve was unclear until the STSCI press conference[8] on January 28. Jonathan Gardner, JWST deputy senior project scientist, described the lightcurve shown in Fig. 8 and explained that for some calibration activities, the space telescope was oriented such as the solar array was pointing at the Sun and the spacecraft was spinning with a 6 hour period. This high variation of brightness could be due to glints on the sunshield.

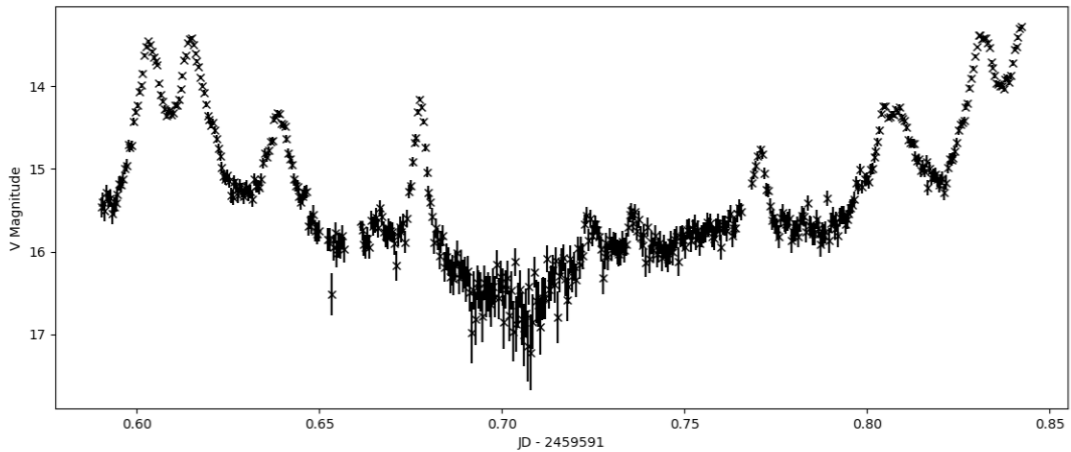

Figure 8. The full lightcurve of JWST extracted from an observation taken on January 12 2022 using the George Mason University 0.8m telescope.

In Figure 9, we show the lightcurve extracted from data obtained by citizen scientist Tateki Goto in Japan on January 8th. From this figure, we can see that all increases in brightness appearing after the double peak are likely caused by the existence of nearby stars inflating the counts in the target aperture. If we ignore these instances, the lightcurve appears flat after the double peak, which is incongruent with what we would expect based on the full lightcurve in Figure 9.

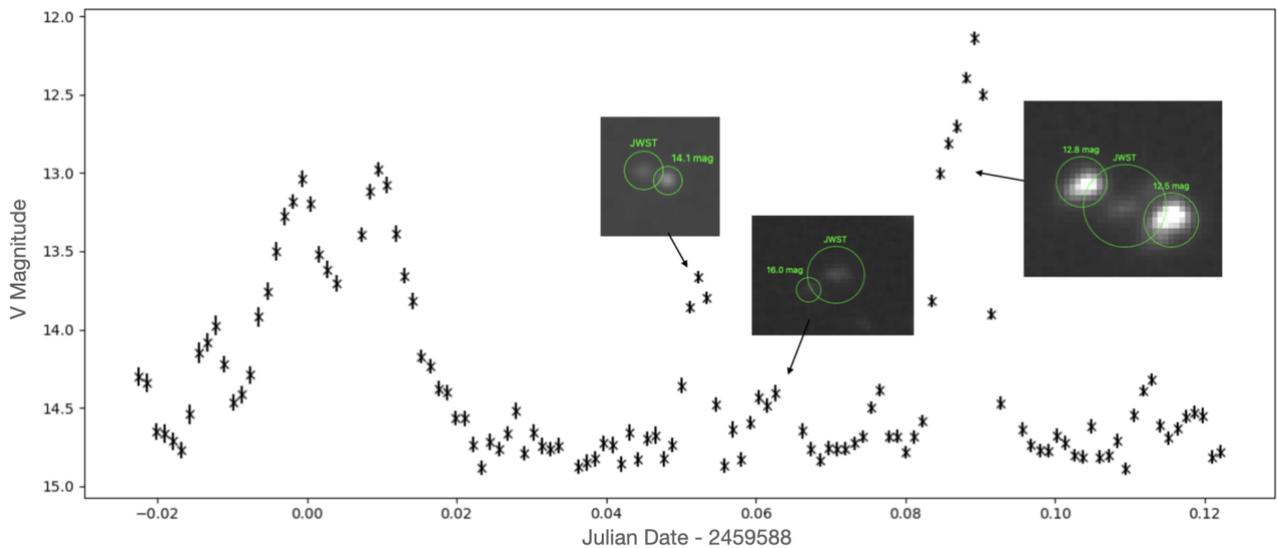

Figure 9. The extracted lightcurve from an observation taken on January 8th. With the exception of the double peak, increases in brightness appear to be caused by nearby stars, suggesting a flat profile beyond the double peak.

We investigated the possibility that the missing peaks are present and hidden by the large increases in brightness from nearby stars. Measuring from the central valley in the double peak, we found that the first peak in brightness is expected to occur after ~0.03 days and the second occurs after ~0.0688 days. In Figure 10, we compare the observation taken on January 12th with that taken on the 8th and find that the times we expect the peaks to occur do not coincide with times where a nearby star interferes with the photometry.

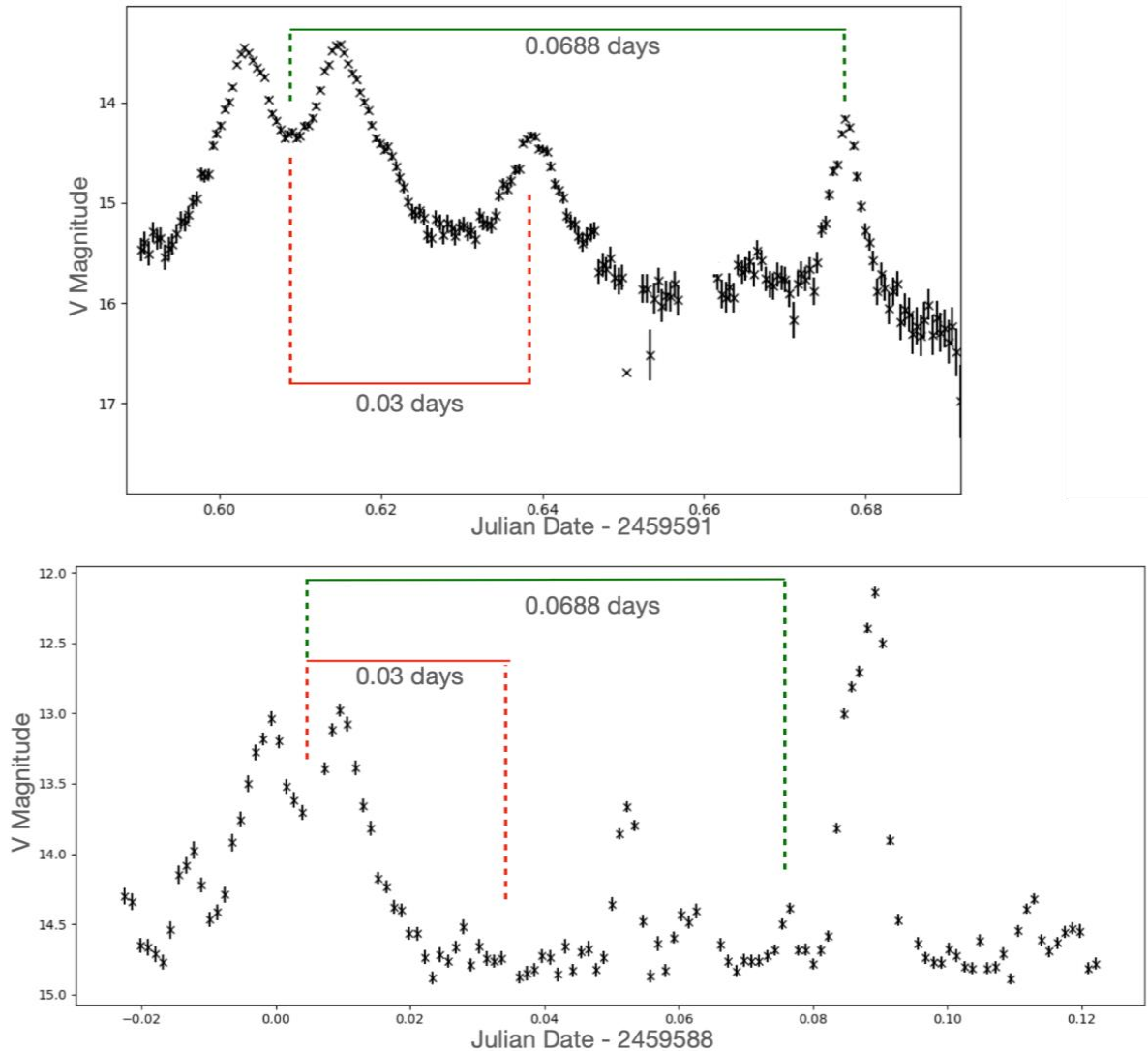

Figure 10. Lightcurves extracted from observations on January 12 2022 (upper) and January 8 2022 (lower). The upper panel is a small portion of the full lightcurve of JWST, observed with the George Mason University 0.8m telescope. The time between the middle of the double peak and the next two peaks to occur is displayed. The lower panel shows a lightcurve extracted from an observation on January 8th by a citizen scientist. At the times when the brightness is expected to increase based on the features seen in the upper panel, this lightcurve stays flat at ~14.6 magnitudes.

Since it does not appear to be the case that the lightcurve's peaks are present but overwhelmed by nearby stars at the time of occurrence, we investigated the possibility that the location of the observation might have affected the appearance of the lightcurve. Figure 11 contains an observation taken on January 6th by the same citizen scientist at the same location as the observation on January 8th. Points in the lightcurve that are contaminated by stars have been indicated. Though this observation does not cover much of the lightcurve after the double peak, it does extend far enough that the peak occurring ~0.03 days after the double peak should be visible. Figure 12 depicts the comparison of the observation of the full lightcurve by Peter Plavchan with that of the observation taken on January 6th. For ease of comparison, the second half of the full lightcurve has been shifted back in time. From this figure, we can confirm the appearance of several features in the full lightcurve that appear in the observation on January6th, including a peak occurring 0.03 days after the double peak. The observation on the 6th (Figures 11 and 12) does not appear to significantly differ from the full lightcurve taken on the 12th, excluding the possibility that the location of the observation on the 8th (Figures 9 and 10) is the cause of the missing peaks in the lightcurve. Further, the peak that occurs 0.03 days after the double peak can be seen in another observation taken on January 13th by citizen scientist Bruno Guillet in Normandy (see subplot 8 of Figure 5), suggesting that location on Earth has very little effect, if any, on the extracted lightcurve.

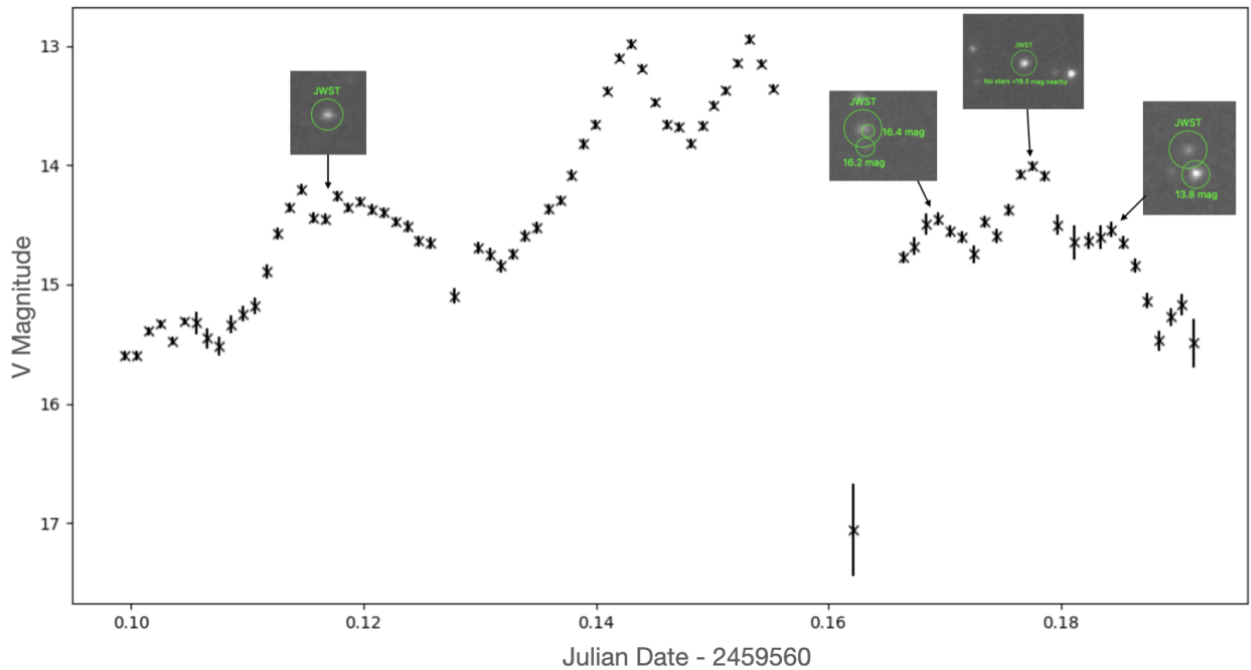

Figure 11. The extracted lightcurve from an observation taken on January 6 2022 from Japan. This observation was taken by the same citizen scientist at the same location as the observation on the 8th with two missing peaks. Areas where stellar contamination is present are indicated.

Finally, it is unlikely that the missing peaks are an artifact of the observation. On January 8th, a second observation occurred at the same time by a different citizen scientist in Japan, Keiichi Fukui. This second observation did not begin early enough for the double peak to be visible in the lightcurve, but it did occur over a timeframe when at least one of the subsequent peaks, occuring after the double peak, should have been visible. This can be seen in Figure 13, which contains the lightcurves of both observations.

We speculate that, in the absence of other compelling arguments for why these peaks are not apparent in observations taken on January 8th, JWST must have been moving in such a way to partially block the reflective material responsible for the peaks in the lightcurve. Based on the timeline of the JWST deployment announced by NASA, it is possible that we are seeing the primary mirror deployment occurring during these observations.

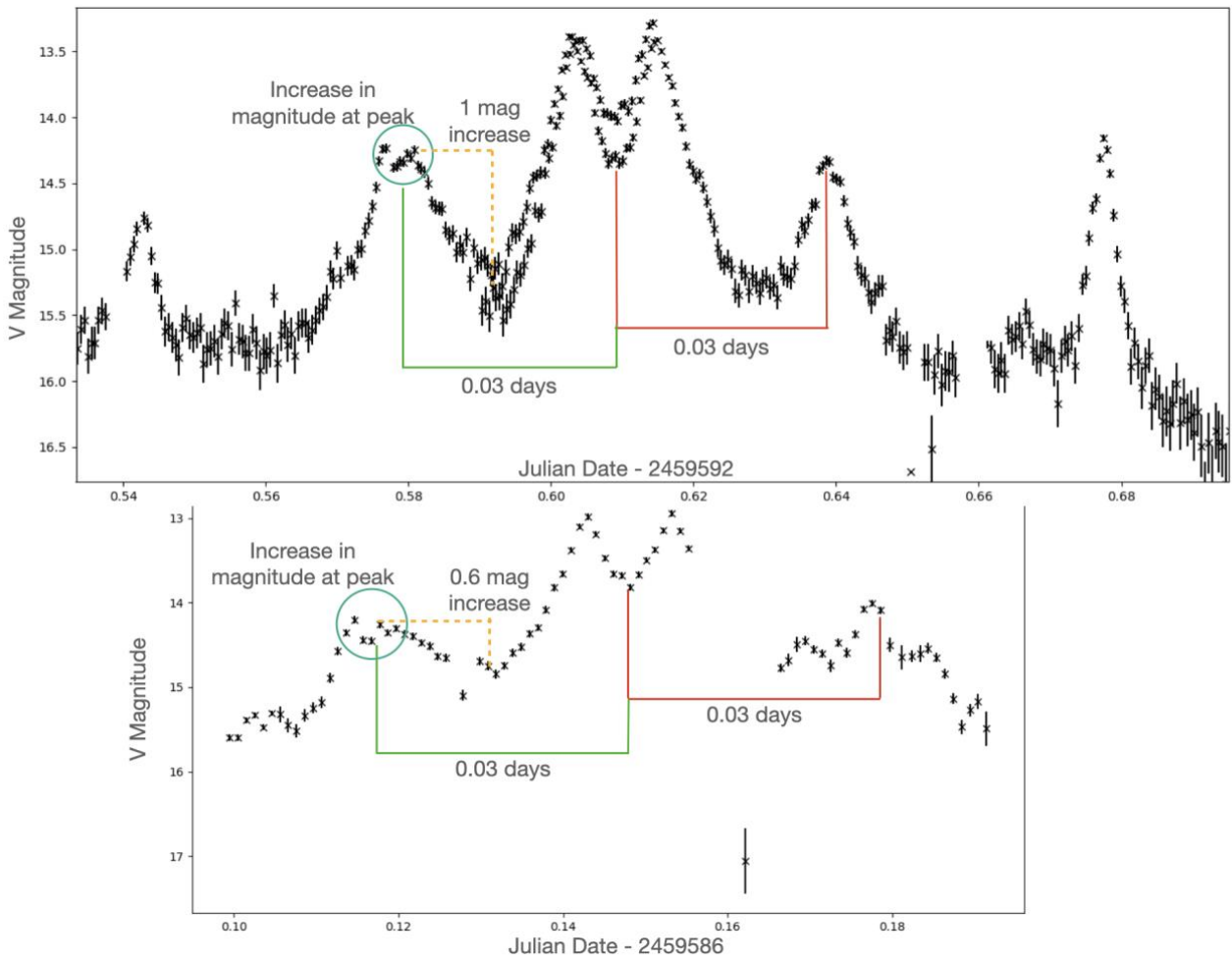

Figure 12. A comparison of features found in the full lightcurve with the observation taken on January 6th from Japan by Tateki Goto. The upper panel is the full lightcurve taken by Peter Plavchan, with features from the second half of the observation shifted back in time for ease of comparison. The lower panel is the extracted lightcurve from the observation taken on the 6th. In both panels the same features can be observed, including a peak 0.03 days after the double peak.

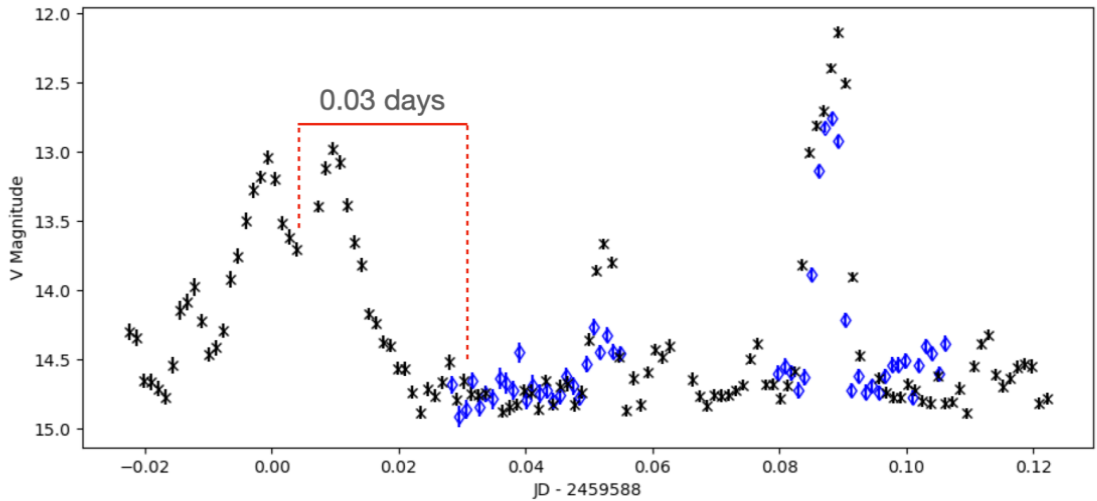

Figure 13. The lightcurve of JWST combining two observations taken concurrently on January 8th. In both observations, the peak that is expected to occur ~0.03 days after the double peak does not appear.

## 4. CONCLUSION

Table 2 summarizes several significant events as JWST drifted toward the second Earth-Sun Lagrange point. Several of those events were witnessed by our network of citizen astronomers distributed around the world. We have detected ambiguously the unfurling of the JWST sunshield, which increased JWST's apparent brightness by at least 1 magnitude. After the deployment, the 6h-lightcurve is complex, displaying a high amplitude and unique features. To date, there is no light curve generator for JWST, or even a tool to simply show the appearance of the spacecraft as seen from Earth available to the scientific community. Consequently, the identification of the source of these features remains difficult.

We have detected a variation in the lightcurve with the transitory disappearance of two peaks that we attributed to a change of geometry due to the deployment of the primary mirror. To confirm the origin of these lightcurve variations, a complex modeling of JWST taking into account the orientation of the space telescope, the source of light, and location of the observers as well as the shape and albedo of the telescope's parts is needed.

Now that the spacecraft has reached the second Earth-Sun Lagrange point, its distance from Earth will remain the same (typically 1.5 million km) and its average magnitude will remain stable as well (V~16.5). This is very close to the limit of detection of the Unistellar eVscope in urban areas (exposure time of 5-10 min). We will likely continue to observe JWST with the citizen scientists, focusing especially on the period of time when JWST is observable for several hours, which will occur in September 2022. Regular observations of JWST and comparison of its lightcurve over time could be useful to reveal the degradation of some parts of the telescope due to micro-impacts and interaction with the solar wind.

This campaign is also an outreach success for the Unistellar network. It is by far the most successful campaign, with 55 observers who collected 145 observations from all the continents except Antarctica. Our simplified ephemeris page was visited approximately 2500 times between December 25 and 30 2021 and has also allowed contributions from other observers around the world who generated the positions of JWST from their location. These numbers are proof of the amateur astronomy community's interest in witnessing astronomical events, including those involving human-made objects.

Despite its relatively small size and low cost, the Unistellar digital telescopes can provide useful information to monitor human-made objects, as well as astronomical events like asteroid flybys[9], occultations[10], transiting exoplanets[11], comets, and more transient events. We are currently developing the scientific applications of our network to allow citizen scientists, schools, and universities to participate in those campaigns and witness activity in space from their home or their colleges. With these networks, we believe that the field of astronomy will become more accessible than ever before.

Table 2. Timeline summary of the JWST launch and deployment as it traveled to L2.

| Date | Event | Days Since Launch |
|---|---|---|
| 2021-12-25 | Launch | 0 |
| 2021-12-25 | Booster and JWST separation detected | 0 |
| 2021-12-26 | Post-MCCM Flare detected by Kendra Sibbernsen | 1 |
| 2021-12-28 | Sunshield deployment begins | 3 |
| 2022-01-03 | Sunshield tensioning complete | 9 |
| 2022-01-04 | Sunshield fully deployed | 10 |
| 2022-01-05 | Secondary mirror deployment complete | 11 |
| 2022-01-06 | First detection of high variation in the lightcurve by Unistellar citizen astronomers | 12 |
| 2022-01-08 | Primary mirror deployment complete (JWST fully deployed at this point) | 14 |
| 2022-01-08 | A peak from the JWST lightcurve does not appear in two observations taken concurrently. These peaks return in subsequent observations. | 14 |
| 2022-01-12 | Beginning of mirror segment deployment | 18 |
| 2022-01-19 | Completion of primary mirror segment deployments (all mirror segments now out of launch position) | 25 |
| 2022-01-24 | JWST reaches L2 | 30 |
| 2022-02-03 | JWST has its first light | 40 |
| 2022-03-11 | Fine phasing alignment completed (mirrors are aligned) | 76 |